\documentclass[amsmath, amssymb, aps, pra,floatfix,nofootinbib,onecolumn,preprintnumbers]{revtex4}
\usepackage[utf8]{inputenc}
\usepackage[letterpaper,top=2cm,bottom=2cm,left=3cm,right=3cm,marginparwidth=1.75cm]{geometry}

\usepackage{tikz}
\usepackage{hyperref}
\usetikzlibrary{quantikz}
\usepackage{amsfonts}
\usepackage{graphicx}
\usepackage{parskip}

\usepackage{graphicx}
\usepackage{amsmath,bm}
\newtheorem{problem}{Problem}
\usepackage[normalem]{ulem}
\usepackage{bbold}
\usepackage{dsfont}

\begin{document}
\title{An exponentially-growing family of universal quantum circuits}

\author{Mo Kordzanganeh}
\author{Pavel Sekatski}
\author{Leonid Fedichkin}
\author{Alexey Melnikov}
\thanks{Corresponding author, e-mail: alexey@melnikov.info
\begin{center}
\fbox{
\begin{minipage}{1\textwidth}
Please check the published version, which includes all the latest additions and corrections: Mach. Learn.: Sci. Technol. 4:035036, 2023, DOI: \href{https://doi.org/10.1088/2632-2153/ace757}{10.1088/2632-2153/ace757}
\end{minipage}
}
\end{center}
}

\affiliation{Terra Quantum AG\\Kornhausstrasse 25, 9000 St.~Gallen, Switzerland}
\begin{abstract}
Quantum machine learning has become an area of growing interest but has certain theoretical and hardware-specific limitations. Notably, the problem of vanishing gradients, or barren plateaus, renders the training impossible for circuits with high qubit counts, imposing a limit on the number of qubits that data scientists can use for solving problems. Independently, angle-embedded supervised quantum neural networks were shown to produce truncated Fourier series with a degree directly dependent on two factors: the depth of the encoding and the number of parallel qubits the encoding applied to. The degree of the Fourier series limits the model expressivity. This work introduces two new architectures whose Fourier degrees grow exponentially: the sequential and parallel exponential quantum machine learning architectures. This is done by efficiently using the available Hilbert space when encoding, increasing the expressivity of the quantum encoding. Therefore, the exponential growth allows staying at the low-qubit limit to create highly expressive circuits avoiding barren plateaus. Practically, parallel exponential architecture was shown to outperform the existing linear architectures by reducing their final mean square error value by up to 44.7\% in a one-dimensional test problem. Furthermore, the feasibility of this technique was also shown on a trapped ion quantum processing unit.
\end{abstract}
\maketitle  
\section{Introduction}

The successes of quantum computing in the past decade have laid the foundations for the interdisciplinary field of quantum machine learning (QML)~\cite{biamonte2017quantum,qml_review_new}, where parameterised quantum circuits (PQC) are used as machine learning models to extract features from the training data. It was argued~\cite{amirapaper} that such quantum neural networks (QNN) could have higher trainability, capacity, and generalization bound than their classical counterparts. Practically, hybrid quantum neural networks (HQNN) have shown promise in small-scale benchmarking tasks~\cite{boston-housing,kordzanganeh2023parallel,schetakis2022review,senokosov2023quantum} and larger-scale industrial tasks~\cite{vw-paper,pharma,rainjonneau2023quantum,kurkin2023forecasting}. Nevertheless, the utility, practicality, and scalability of pure QNNs are still unclear. Furthermore,~\cite{schuld-advantage} provided a thorough overview of this field, showing that while classical machine learning is solving large real-world problems, QNNs are mostly tried on synthetic, clean datasets and show no immediate real-world advantages in its current state\footnote{This is true for pure quantum models trained on classical datasets. There are numerous successes in applying quantum models to quantum-native problems~\cite{hhl,vqe,discocat}.}. It also suggested that the QNN research focus should be shifted from seeking quantum advantage to new research questions, such as finding a new, advantageous quantum neuron. This work explores a new quantum neuron that argues for moving beyond using the Pauli-, single-qubit gates for encoding data and instead employing higher-dimensional unitaries through gate decomposition. 

Since quantum gates are represented by elements of compact groups, Fourier analysis is a natural tool for analysing QNNs. \cite{schuld_fourier} showed that certain quantum encodings can create an asymptotically universal Fourier estimator. A universal Fourier estimator is a model that can fit a trigonometric series on any given function. As the number of terms in the series approaches infinity, the fit becomes an asymptotically perfect estimate. This estimator can initially infer the coarse correlations in the supplied data, and by increasing the number of Fourier terms, it can incrementally judge the more granular properties of the dataset. This provides an adjustable, highly-expressive machine learning model. 

In \cite{schuld_fourier}, the authors showed that a QNN could operate as such in two ways: a \emph{sequential} single-qubit architecture with $n$ repetitions of the same encoding could yield $n$ Fourier bases, which could also arise from an $n$-qubit architecture with the encoding gates applied to all qubits in \emph{parallel}.

The sequential architecture is widely shown to be an efficient Fourier estimator, demonstrated both theoretically \cite{schuld_fourier} and empirically~\cite{data_reuploading,mo_paper}. However, sequential circuits are often deep \cite{mo_paper}, and assuming that near-term quantum computers will have a non-negligible noise associated with each gate, these circuits can experience noise-induced barren plateaus~\cite{noise-bp}. Barren plateaus are a phenomenon observed in optimization problems where the gradients of the model vanish, rendering them impossible to train. More importantly, a single qubit can be simulated efficiently in a classical setting, so this architecture brings no quantum advantage.

In contrast, the parallel setting offers the exponential space advantage of quantum computing but poses two challenges for large numbers of qubits:  
\begin{itemize}
    \item An exponentially growing parameter count with the number of qubits required to span the entire group space $\mathcal{SU}(2^n)$~\cite{haar_measure_paper}, which could also lead to noise-induced barren plateaus. Spanning the full space is especially important for a priori problems, where the best model lies in the Hilbert space. Still, the machine learning scientist has no prior knowledge of parameterising a circuit to reach this point. In addition, gradient calculations in QNNs -- as of this publication -- use the parameter-shift rule discovered and presented in \cite{parameter-shift}. This method requires two evaluations of the QNN to find the derivative of the circuit with respect to each of the trainable parameters. An exponentially growing number of trainable parameters translates to exponentially increasing resources required for gradient computation. In mitigation, \cite{poly1,poly2} showed that a polynomially-growing number of parameters could generate a similar result based on the quantum $t$-design limits~\cite{t-design}.
  
   \item Strongly-parameterised QNNs have a vanishing variance of gradient which decreases exponentially with the number of qubits~\cite{bp}. This means that for a large number of qubits if one initialises her QNN randomly, they will encounter a barren plateau. This happens because the expectation value of the derivative of the loss function with respect to each variable for any well-parameterised quantum circuit is zero, and its variance decreases exponentially with the number of qubits. Mitigation methods have been suggested in \cite{bp-mit1,bp-mit2}, and most notably in \cite{log-nobp}, it was shown that by relaxing the well-parameterised constrain to only include a logarithmically growing circuit depth with the number of qubits in the system and use local measurements, the circuit is guaranteed to evade barren plateaus. \cite{bp-zx} developed a platform based on ZX-calculus\footnote{A graphical language, based on tensor networks, to analyze quantum circuits\cite{zx1}.} to explore which QNN architectures are affected by the barren plateau phenomenon and found that strongly-entangling, hardware efficient circuits suffer from them. In contrast with the previously-mentioned cases of barren plateaus, the latter is not noise-induced. Thus, this problem must be addressed even in the fault-tolerant future of quantum computing.
\end{itemize} 

Therefore, the practising QML scientist is limited in choosing her QNN architectures for general data science problems: they need to be shallow\footnote{\cite{exp_inc_1,exp_inc_2} show that even in large, shallow circuits, the trainability can be sub-optimal. They claim that the landscape of solutions of shallow circuits has an exponentially increasing number of undesirable local minima.} or employ only a few qubits. This contribution suggests modifying the encoding strategies in \cite{schuld_fourier} to increase the growth of the Fourier bases in a QNN from linear in the number of qubits/number of repetitions to exponential. The proposed encoding is constructed by decomposing large unitary generators into local Pauli-Z rotations. This improves the expressivity of the QNNs without requiring additional qubits or encoding repetitions. The increased expressivity is a product of eliminating the encoding degeneracies of the quantum kernel, making efficient use of the available Hilbert space by assigning a unique wavevector to each of its dimensions. However, such encodings could introduce a greater risk of limiting the model's Fourier accessibility\footnote{Each of these bases has a Fourier amplitude and a phase angle, which need to be altered to fit a model on the training data. However, depending on the architecture (of both the encoding layers and the trainable layers), these quantities may be limited in the values they can represent. This could significantly limit their ability to represent various functions. This will henceforth be referred to as the \emph{Fourier accessibility} of the quantum architectures.}. 


Sec.~\ref{sec:lin_arch} provides a review of how angle-embedded QNNs approximate their input distributions by fitting to them a truncated Fourier. Specifically, Sec.~\ref{sec:lin_arch} reviews the linear encoding architectures and how their number of Fourier bases grow linearly with the number of repetitions -- sequential linear in Sec.~\ref{sec:seq_lin} -- as well as the number of qubits -- parallel linear in Sec.~\ref{sec:par_lin}. Then, Sec.~\ref{sec:exp_arch} introduces the same two architectures but slightly modified to represent an exponentially-growing number of Fourier bases. To use these architectures in practice, Sec.~\ref{sec:training} compares the training performance of these architectures and shows that the parallel exponential has a superior training performance to the other architectures on a synthetic, one-dimensional dataset. Finally, Sec.~\ref{sec:critical} critically evaluates the work and suggests areas for future investigation.


\section{Background review -- linear architectures} \label{sec:lin_arch}

As discussed in \cite{schuld_fourier}, all quantum neural networks that use angle embedding\footnote{Ref.~\cite{schuld_kernel} explores other embedding strategies, too, such as basis embedding and amplitude embedding. This work primarily focuses on angle embedding. Nevertheless, it is worth noting that in the circuit model, all circuit parameters enter as angles at some level of the description.} as their encoding strategy produces a truncated Fourier series approximation to the dataset. 
\cite{schuld_fourier} also specifically explored two families of architectures of quantum neurons: a single-qubit architecture with a series of sequential $\mathcal{SU}(2)$ gates and a multi-qubit architecture with parallel $\mathcal{SU}(2)$ encoding gates. In this section, the results and the architectures introduced in \cite{schuld_fourier} are explored in depth in Sec.~\ref{sec:exp_arch}. Two alternative QNN architectures are presented with the capability of achieving an exponentially higher Fourier expressivity for the same number of gates.

Consider a quantum neuron that maps a real feature $x \in \mathcal{X}$ onto the quantum circuit via a parametric gate $S(x)=e^{-i\mathcal{G}x}$. In most common architectures, the only parametric gates are single qubit rotations $\{R_x,R_y,R_z\}$. For this work, the Pauli-Z generated rotations are used without any loss of generality $\mathcal{G}=\frac{1}{2}\sigma_z=\frac{1}{2}\big(\begin{smallmatrix}
  1 & 0\\
  0 & -1
\end{smallmatrix}\big)$\footnote{extra gates to convert between different generators can be absorbed into the variational gates -- see \cite{schuld_fourier}}, then the embedding gate takes a simple form $S(x)= \big(\begin{smallmatrix}
 e^{-i x /2 } & 0\\
  0 & e^{i x /2}
\end{smallmatrix}\big)$. In general, the dependence of the expected value of any observable on the parameter $x$ is then given by
\begin{equation*}
    \langle M\rangle = \bra{\Psi}(S^\dag(x)\otimes \mathbb{1}) M (S(x)\otimes \mathbb{1})\ket{\Psi} = (c_0+c_0^*) + c_{1} e^{i x}+ c_1^* e^{-i x}
\end{equation*}
with some complex parameters $c_0$ and $c_1$, which depend on the rest of the circuit and the measurements.

This expected value is a function of the feature $x$ with a very simple Fourier series. The \textit{data re-uploading method}~\cite{data_reuploading} is a natural way to construct neurons that give rise to richer Fourier series. These are architectures where several parametric gates depend on the same $x$. It is the most straightforward to consider gates that have a hardwired dependence on the feature\footnote{In principle, one may also consider gates $S(x;\theta)= e^{-i f_\theta(x) \mathcal{G}}$ with a \textit{variational} dependence on the feature $x$, where the parameters $f_\theta$ are to be learned. However, this could present additional challenges related to the computation of gradients and increased number of parameters}. In particular, such that the expected value of any observable takes the form of a discrete Fourier series 
\begin{equation}
\label{eqn:fourier_general}
    f(x,\theta) =\sum_{k} c_k(\theta) e^{ikx},
\end{equation}
where $\theta$ the variational parameters and  $c_k \in \mathbb{C}$ with $c^{*}_{-k} = c_k$ for real observables. In Sec.~\ref{sec:seq_lin} and \ref{sec:par_lin}, two architectures exhibited in \cite{schuld_fourier} are reviewed. The \textit{Fourier expressivities} of these architectures are of particular interest, that is, the list of wavenumbers $\{k_1,k_2,\dots\}$ appearing in the exponents in Eq.~\eqref{eqn:fourier_general}.

\subsection{Sequential linear}\label{sec:seq_lin}

\begin{figure*}
     \centering
     \includegraphics[width=1\linewidth]{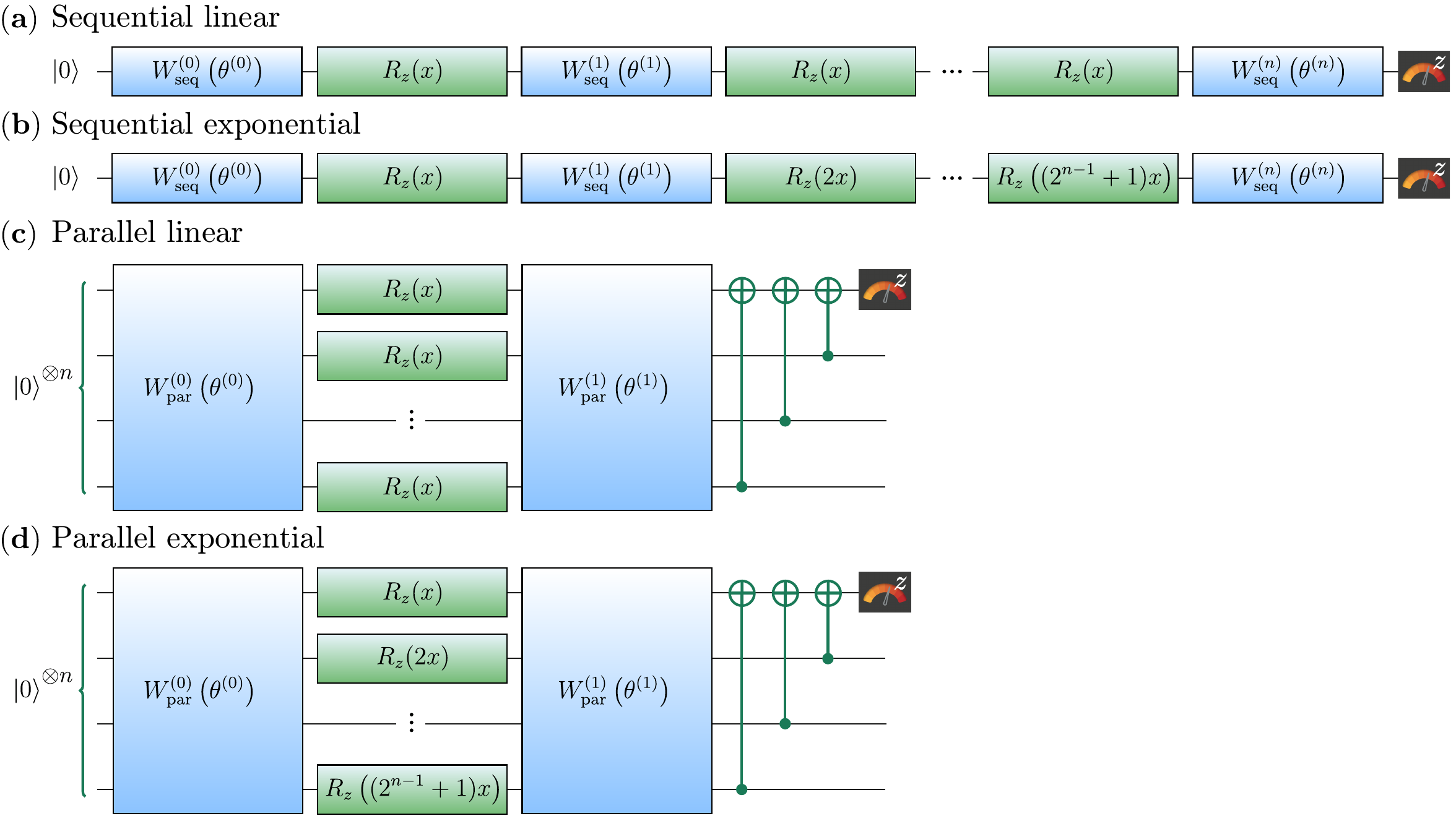}
        \caption{The four general circuits under analysis in this paper. In the exponential architectures, the first encoding is kept the same, and the subsequent encoding gates are multiplied by the coefficients in Eq.~(\ref{eqn:numbers}). The parallel circuits have a CNOT layer at the end to ensure that all qubits are cooperating in the training by propagating the $\pi$-measurement through all quantum wires.}
\label{fig:archs}
\end{figure*}

The single-qubit sequential linear method uses repetitions of the same single-qubit encoding gate $S(x)$ interlaced in-between trainable variational layers. Fig.~\ref{fig:archs}a shows this implementation with generalized variational gates. Since the eigenvalues of each unitary are $e^{\pm i \frac{1}{2}x}$, it is
straightforward to observe (see, e.g. App.~\ref{appendix:seq_lin}) that after $n$ encoding layers, the expected value of any observable takes the form
\begin{equation}
\label{eqn:fourier_first}
    f(x,\theta) =\sum_{k=-n}^{n} c_k(\theta) e^{ikx},
\end{equation}
Thus, the repetitions have an additive effect such that for $n$ repetitions, the final list becomes $\{-n,-n+1,\cdots,0,\cdots,n-1,n\}$. Each of the wavenumbers in the list gives rise to a sinusoidal term with the same frequency. Therefore, for $n$ repetitions of the encoding $S(x)= e^{-i\mathcal{G}x}$, $n$ distinct Fourier bases are generated.

\subsection{Parallel linear}\label{sec:par_lin}
In the parallel setting, the single-qubit encoding gates are applied in parallel on separate qubits -- see Fig.~\ref{fig:archs}c. Similarly to the sequential encoding, for $n$ parallel rotations $n$ Fourier bases are produced. This is due to the commutativity between the parallel rotations as they act on separate qubits. The generator $\mathcal{G}$ becomes:
\begin{equation}
    \mathcal{G} =  \frac{1}{2} \sum_{q=1}^{r} \sigma_z^{(q)},
\end{equation}
where $q$ is the qubit index, $r$ indicates the total number of qubits, and $\sigma_z^{(q)}$ is the Pauli-Z matrix applied to the $q^\text{th}$ qubit. In App.~\ref{appendix:par_lin_proof}, it is shown that $\mathcal{G}$ -- being a square matrix of dimensions $2^r$ -- has $2r+1$ unique eigenvalues. This suggests a high degree of degeneracy in its eigenspectrum. As before, subtracting these values from themselves yields a list of wavenumbers ranging from $-r$ to $r$ generating $r$ Fourier bases.

\subsection{Redundancy}

Both in the sequential and parallel linear architectures, there is a lot of redundancy in how the feature is encoded into the circuit. This is the easiest to see for the parallel architecture, where most of the eigenvalues of $\exp(i x \mathcal{G})$ are largely degenerate as the encoding commutes with qubit permutations.

\section{Results -- exponential architectures}\label{sec:exp_arch}

In this section, two new families of architectures are suggested that can encode an exponential number of Fourier bases for a given number of repetitions/parallel encodings. The basis of this generalization is to modify each "subsequent" appearance of the encoding gate in the circuit by a re-scaling of the generator $S(x)\to S(m x)$ with an integer $m$. Keeping the factors $m$ integer guarantees that this procedure results in a discrete Fourier series in the form of Eq.~\eqref{eqn:fourier_general}.

\subsection{Sequential exponential}\label{sec:seq_exp}
It was shown in Sec.~\ref{sec:seq_lin} that the wavenumbers created in the linear models are highly degenerate. By modifying the circuit encoding, this degeneracy can be reduced, resulting in adding new wavenumbers to the list. This is accomplished by altering the generators in the individual encoding layers. In the linear case, the diagonal elements of the generator $\lambda_i$ always belonged to the list $\{-\frac{1}{2},\frac{1}{2}\}$, but could be altered by scaling the generator $\mathcal{G}$ in each layer. In practice, this is achieved by scaling the embedded data $x$ and mathematically associating it with the generator. The resultant function becomes

\begin{eqnarray*}
    f(x,\theta) =\left( W_{1,j_1}^{(0)\dagger}W_{j_1,j_2}^{(1)\dagger}\cdots M_{k',k} \cdots W_{i_2,i_1}^{(1)} W_{i_1,1}^{(0)}\right)\\
e^{\left((\lambda^{(1)}_{j_1}+\lambda^{(2)}_{j_2}+\cdots)-(\lambda^{(1)}_{i_1}+\lambda^{(2)}_{i_2}+\cdots)\right)x},
\end{eqnarray*}

where $\lambda^{(l)}_i = a_l \lambda_i = \frac{1}{2}\{-a_l,a_l\}$ for $a_l \in \mathbb{N}$~\footnote{In comparison, for linear architectures $a_l=1$ for all $l$.}. In this work, $a_l$ scales as follows $a_l =\{2^0,2^1,2^2,\cdots,2^{n-1}+1\}$. The motivation behind this choice is the sum of powers of 2, $\sum_{i=0}^{n-1} 2^i = 2^{n} - 1$, where the largest wavenumber possible, $2^n$, is obtained by taking all the positive contributions from the list of eigenvalues, i.e. $k_{max}= \sum_{i=0}^{n-1} 2^i + 2^{n-1} + 1 = 2^n$. Next, one can switch the signs of the positive values to negative starting from the smallest term to produce all integers from $-2^n$ to $2^n$. This generates $2^n$ Fourier frequencies. Fig.~\ref{fig:archs}b shows a quantum circuit encoded using the sequential exponential strategy with 2 layers. App.~\ref{appendix:constrained} demonstrates that this network produces extreme constraints on the Fourier accessibility and thus is an undesirable choice for general data modelling. However, this scheme motivates extending this idea to parallel architectures.

\begin{figure*}[t]
    \centering
    \includegraphics[width=\linewidth]{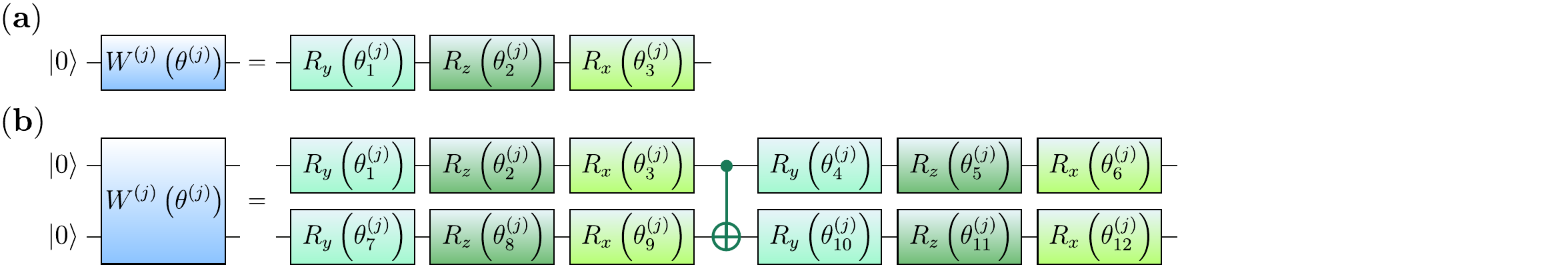}
    \caption{Variational gates used in single -- (a) and two-qubit (b) experiments.}
    \label{fig2:variational}
\end{figure*}

\subsection{Parallel exponential}\label{sec:exp_par}
To perform this extension, it is appropriate to proceed with a two-qubit example. The parallel linear method described in Sec.~\ref{sec:par_lin} produces the generator:
\begin{equation}
\mathcal{G^{\text{lin}}}=\frac{1}{2} \left(\sigma_z \otimes \mathbb{I} + \mathbb{I} \otimes \sigma_z\right) = \begin{pmatrix}
  1 & 0 & 0 & 0\\
  0 & 0 & 0 & 0\\
  0 & 0 & 0 & 0\\
  0 & 0 & 0 & -1
\end{pmatrix}
\end{equation}
This matrix has three unique eigenvalues, $\lambda \in \{-1,0,1\}$, and when subtracted from itself -- yielding wavenumbers $L^{\text{(lin)}}_k = \{-2,-1,0,1,2\}$ -- it can produce 2 Fourier bases with frequencies $\{1,2\}$. One could generate a matrix with more unique values. For example,
\begin{equation}
\mathcal{G^{\text{exp}}}=\frac{1}{2} \left(3\sigma_z \otimes \mathbb{I} + \mathbb{I} \otimes  \sigma_z\right)=\begin{pmatrix}
  2 & 0 & 0 & 0\\
  0 & -1 & 0 & 0\\
  0 & 0 & 1 & 0\\
  0 & 0 & 0 & -2
\end{pmatrix},\label{eqn:exp_matrix}
\end{equation}
is a generator with four unique eigenvalues that generate nine wavenumbers $\{-4, -3, -2, -1, 0, 1, 2, 3, 4\}$. This generator can be constructed using the quantum circuit shown in Fig.~\ref{fig:archs}d. In this case, a $\mathcal{SU}(4)$ generator is employed. This is decomposed into two $\mathcal{SU}(2)$ generators, one using the group parameter, $x$, and the other $3x$. This can be generalized to $n$ qubits as one can extend the matrix for larger numbers of qubits, i.e. for $n$ qubits $\mathcal{G}$ would be a diagonal matrix starting from $-2^{(n-1)}$ up to $2^{(n-1)}$, producing $2^n$ Fourier bases. The quantum circuit associated with this generator is an application of Pauli-Z rotations of $x$ with frequencies increasing in the following way: \begin{equation}\label{eqn:numbers}
   \mathcal{L} = \{2^0,2^1,2^2,...,2^{n-1}+1\},
\end{equation}where $n$ is the number of qubits. Note the similarities between the sequential and parallel encodings and their symmetries in how the circuits are constructed. One also recognizes similarities between the parallel encoding and Kitaev's quantum phase estimation algorithm~\cite{kitaev1995quantum}, albeit in this case, $x$ is a classical feature.

This can be significantly more expressive than the parallel linear method. Still, this advantage needs to be accompanied by Fourier accessibility. If the Fourier values of these newly-acquired bases cannot be altered, there would be no advantage in pursuing this setting. Sec.~\ref{sec:training} shows a significant advantage in using parallel exponential encoding in a simple toy example.

\begin{figure*}[t]
    \centering
    \includegraphics[width=\linewidth]{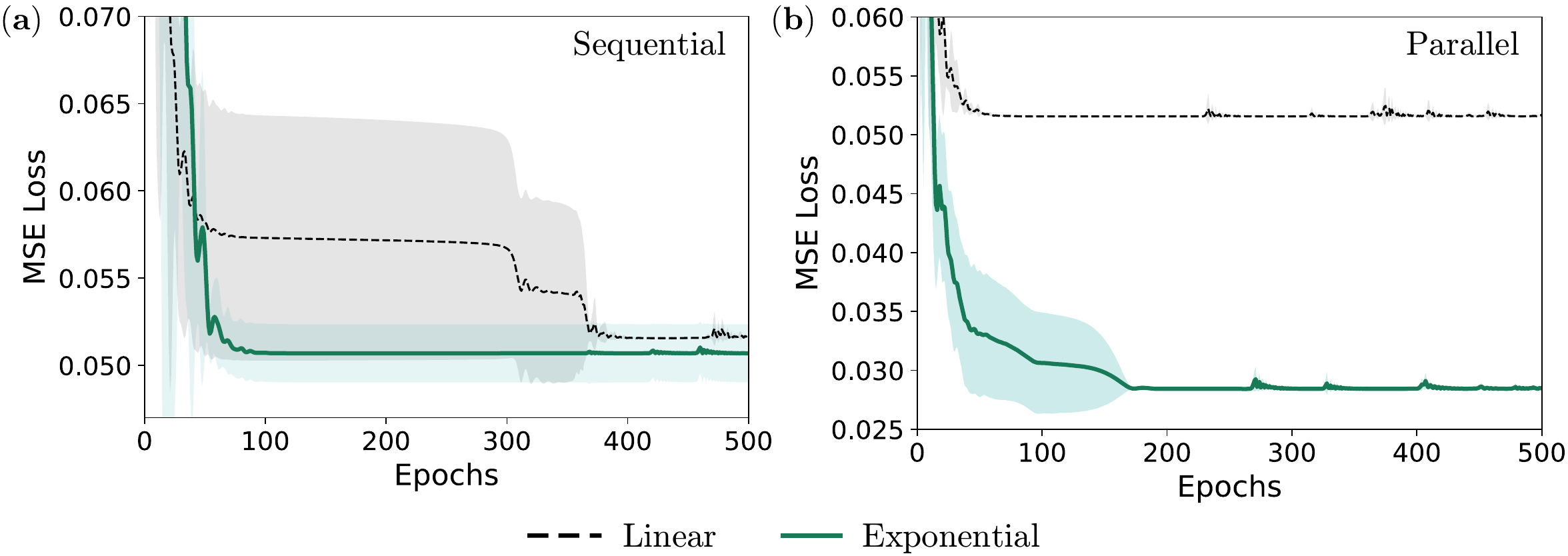}
    \caption{Training losses indicate a training advantage for the parallel exponential, and the sequential exponential architecture performs only marginally better than the linear architectures. The training was done on QMware hardware~\cite{qmware} using the PennyLane Python package~\cite{pennylane}. The Adam optimizer minimises a mean squared loss function with a learning rate of $\epsilon=0.1$ and with uniformly-distributed parameters $\theta \in [0,2\pi]$.}
    \label{fig3:losses}
\end{figure*}

\begin{figure*}[t]
    \centering
    \includegraphics[width=\linewidth]{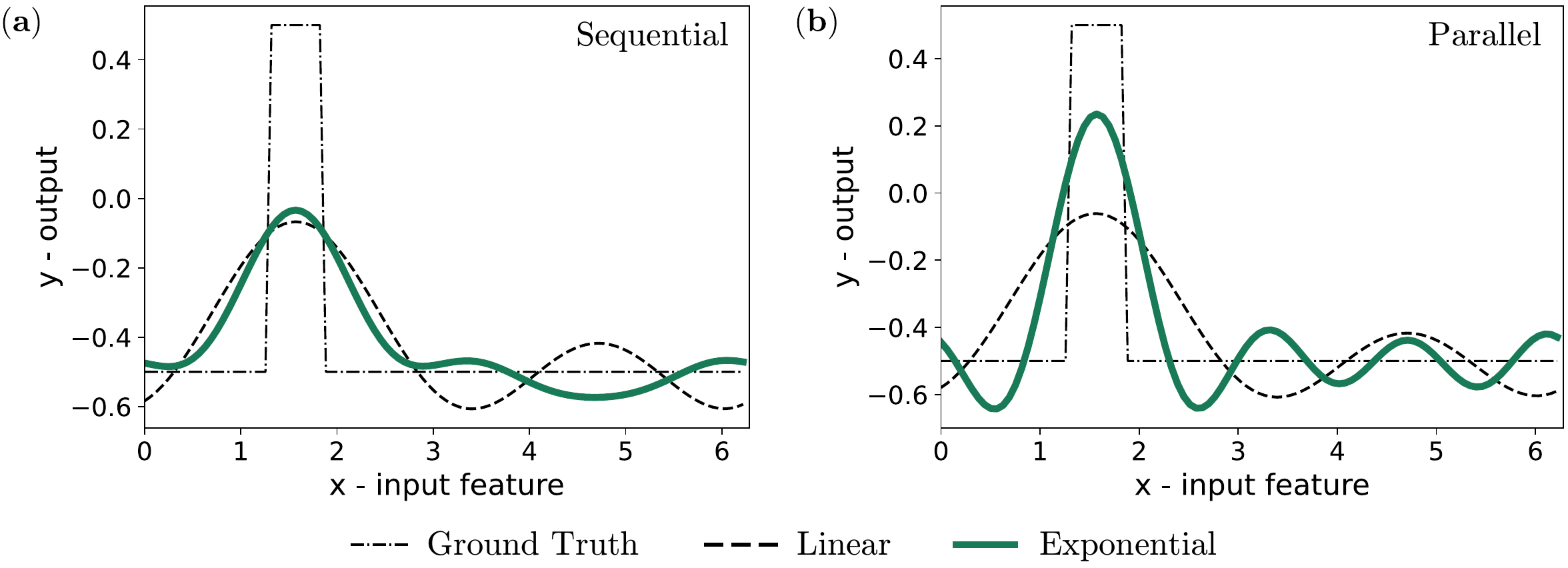}
    \caption{The QNNs fit the best possible truncated Fourier series on the top-hat function. The parallel exponential architecture provides the best fit. Even though the sequential exponential architecture has access to the same four Fourier frequencies, it fails to access all of them efficiently, and as a result, it performs sub-optimally. The linear architectures perform similarly to each other, potentially arising from their high Fourier accessibility to the two Fourier frequencies that they can represent.}
    \label{fig4:tophat}
\end{figure*}

\begin{figure*}
    \centering
    \includegraphics[width=\textwidth]{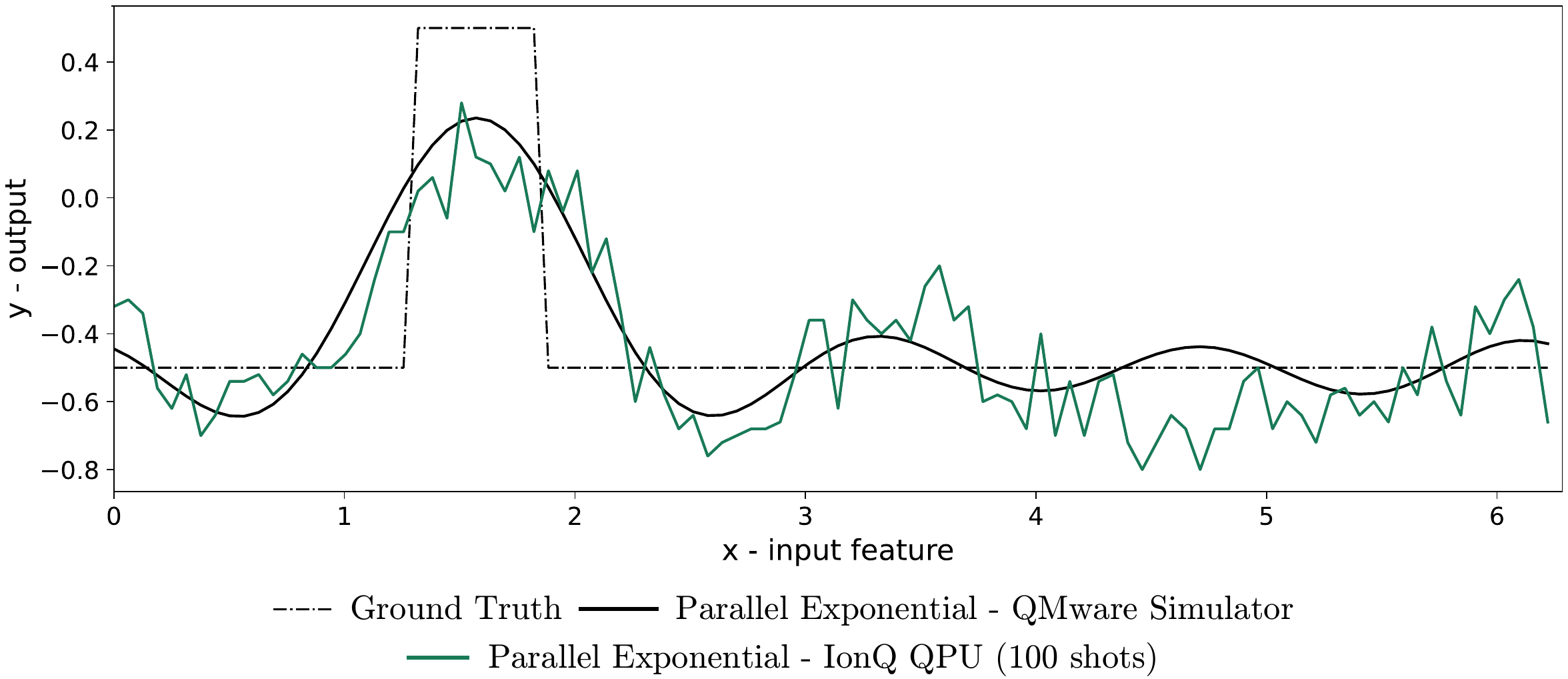}
    \caption{The parallel exponential fit to the top-hat function on a simulator vs on the IonQ Harmony quantum processor. The noisy solid line evaluates this network for 100 equally-spaced points using 100 shots of this device.}
    \label{fig5:qpu}
\end{figure*}

\subsection{Training}\label{sec:training}
\begin{figure*}[ht]
     \centering
     \includegraphics[width=\linewidth]{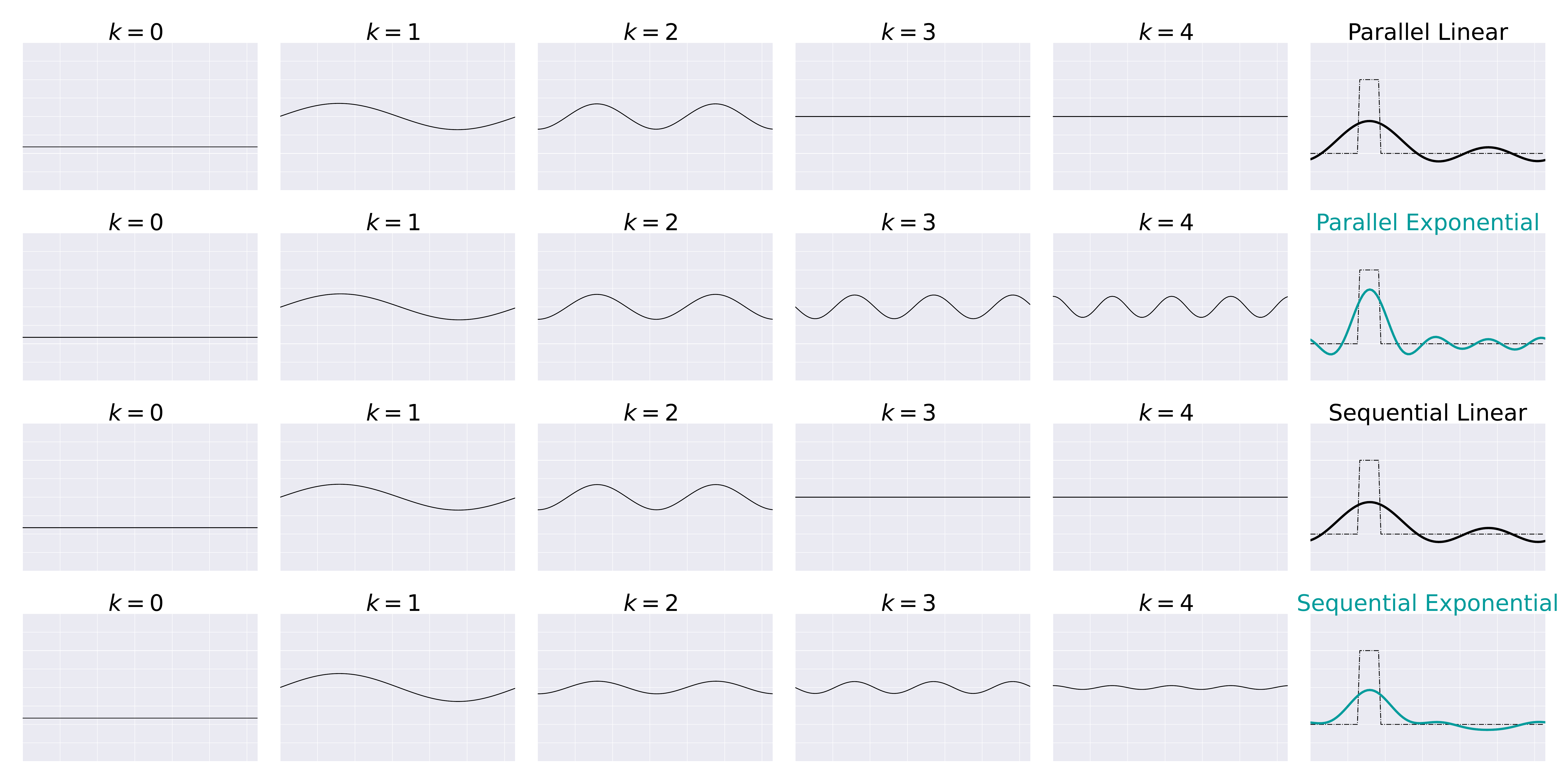}
        \caption{Fourier decomposition of the four architectures after training to fit the top-hat function. The linear architectures can only access two Fourier frequencies, whereas the exponential ones can access four.}
        \label{fig6:decomposition}
\end{figure*}

In this section, the training performance of these four architectures on a simple dataset is compared. Each architecture is trained to reproduce a one-dimensional top-hat function. Fig.~\ref{fig4:tophat} shows the ground truth, as well as the fitting performances of these architectures, and Fig.~\ref{fig3:losses}, shows their training performance. It is clear that the parallel exponential architecture fits a closer function to the ground truth, and in contrast, the sequential exponential architecture has the worst performance of all models. Furthermore, the Fourier decompositions of the models in Fig.~\ref{fig6:decomposition} show that exactly two Fourier terms are accessed by the linear architectures and four by the exponential ones. Additionally, Fig.~\ref{fig5:qpu} demonstrates the performance of the parallel exponential architecture on a trapped ion quantum processor. The IonQ implements a high-fidelity gate-based quantum processing unit through a process known as laser pumping trapped-ions explained in~\cite{ionq-paper}. The hardware was shown to be one of the most accurate in recent benchmarking tests~\cite{benchmarkingQMW}. We specifically used the hardware introduced in~\cite{11qubit-ionq} with a single-qubit fidelity of $0.997$ and a two-qubit fidelity of $0.9725$. The code implementation was done through Amazon Web Services (AWS) Braket, and the process of the forward pass for 100 data points took four hours and 11 minutes due to the delays and queuing times. It can be observed that the low number of shots is the dominant source of noise here, and higher shot counts could yield a smoother curve that is closer to the simulator. It is noteworthy that even though the superconductor-based QPUs were not tested here, they are expected to produce a similar result if they match the single- and two-qubit gate fidelity rates.  

\subsection{Critical Evaluation}\label{sec:critical}
While the results for the parallel exponentials are encouraging, it is equally important to understand the limitations of this approach. Firstly, while the exponential growth in the number of Fourier frequencies is evident, this is not the higher limit of Fourier frequency growth. \cite{schuld_fourier} showed that for $L$ repetitions of an encoding gate with a Hilbert space of dimension $d$, there is an upper limit to this growth of the form

\begin{equation}
K < \frac{d^{2L}}{2} -1,
\end{equation}
where $K$ is the number of Fourier frequencies. This suggests a potential for square-exponential growth, whereas the method discussed in this work only grows exponentially. In App.~\ref{appendix:maths_problem}, a mathematical problem is proposed whose solution could unlock the maximum possible Fourier accessibility. 

Secondly, it is important to emphasize that the two parallel architectures are the same, with a minor multiplicative factor added in the exponential case. Training them for a fixed number of epochs requires the same computational resources. However, adding more Fourier bases by eliminating the network's degeneracy could result in under-parameterised models. Therefore, it is often necessary to parameterise the exponential architectures more heavily than the linear ones, indirectly affecting the required resources. Every Fourier frequency requires two degrees of freedom (real-valued parameters), and an exponentially-growing Fourier space requires the resources to grow exponentially, too. These resources could include the classical memory required to store the parameters or the classical optimizer that needs to calculate the gradient for these parameters. And lastly, extending this to many qubits will still result in barren plateaus.

\section{Conclusion and future work}\label{sec:conclusion}
This work suggested two new families of QNN architectures, dubbed sequential and parallel exponential circuits that provided an exponentially growing Fourier space. It was demonstrated that the former struggled with accessing these frequencies but also that the latter showed an advantage in approximating a top-hat function. 

Future work could focus on a quantitative understanding of the Fourier accessibility of these networks, such that the optimal variational parameterisation could be chosen for a specific problem. Another possible direction for future work is to depart from hardwired encoding gates. A natural elementary step in this direction is to consider single-qubit gates of the form $S_i(x,w_i)= \exp(- i \, x w_i \frac{1}{2}\sigma_z )$, where the scaling factor $w_i$ is an independent scalar trainable parameter for each occurrence of the encoding gate in the circuit. In this case, the final wavevectors $k$ are linear combinations of the parameters $w_i$ that can be potentially trained efficiently. As an added note, the parallel exponential encoding introduced in this work for up to two qubits coincides with the commendable work in \cite{shinetal}. This paper came to our attention after we had released the preprint, and we recognise that the parallel exponential architecture bears resemblance to the Trenary encoding both for the two-qubit case and in the type of growth in Fourier terms, albeit with different scaling strategies. Furthermore, \cite{fockstateenhanced} follows a similar example and creates this architecture for an optical setup. 

\bibliography{ref}
\bibliographystyle{unsrt}

\clearpage

\appendix
\section{Fourier estimation of the linear architectures}

\subsection{Sequential linear} \label{appendix:seq_lin}
To see the extraction of the Fourier modes more explicitly, one can write the function represented by this circuit as the expectation value of a measurement operator $M$:
\begin{equation}
    f(x,\theta) = \bra{0}W^{(0)\dagger}(\theta)S^{\dagger}(x)W^{(1)\dagger}(\theta)\cdots M \cdots W^{(1)}(\theta) S(x) W^{(0)}(\theta)\ket{0}
\end{equation}
\begin{equation}
     = \bra{0}W^{(0)\dagger}(\theta)e^{i\mathcal{G}x}W^{(1)\dagger}(\theta)\cdots M \cdots W^{(1)}(\theta) e^{-i\mathcal{G}x} W^{(0)}(\theta)\ket{0}
\end{equation}
Writing this in tensor notation yields 
\begin{equation}
     = W_{1,j_1}^{(0)\dagger}e^{i\mathcal{G}_{j_1,j_2}x}W_{j_2,j_3}^{(1)\dagger}\cdots M_{k',k} \cdots W_{i_3,i_2}^{(1)} e^{-i\mathcal{G}_{i_2,i_1}x} W_{i_1,1}^{(0)},
\end{equation}
where $W = W(\theta)$ and the index summation convention is employed. We can simplify this expression further by employing $$e^{-i\mathcal{G}_{i_2,i_1}x} = e^{\lambda_{i_1} x} \delta_{i_1,i_2},$$ where $\lambda_{i_1} = \{-\frac{1}{2},\frac{1}{2}\}$ represents the eigenvalues of $\mathcal{G}$. This simplifies our function:
\begin{equation}\label{eqn:exps_out}
    =\left( W_{1,j_1}^{(0)\dagger}W_{j_1,j_2}^{(1)\dagger}\cdots M_{k',k} \cdots W_{i_2,i_1}^{(1)} W_{i_1,1}^{(0)}\right) e^{\left((\lambda_{j_1}+\lambda_{j_2}+\cdots)-(\lambda_{i_1}+\lambda_{i_2}+\cdots)\right)x}.
\end{equation}
This means that for every combination of $\{\{i_1,\cdots\},\{j_1,\cdots\}\}$, the QNN produces a wave-front with a wavenumber\footnote{Note that this happens at the level of the density matrices and is independent of the choice of $M$.} $k_{\boldsymbol{i},\boldsymbol{j}} = (\lambda_{j_1}+\lambda_{j_2}+\cdots)-(\lambda_{i_1}+\lambda_{i_2}+\cdots),$ where $\boldsymbol{i},\boldsymbol{j} \in \mathbb{N}^n$. Note that for $n$ repetitions of this encoding, one obtains $2^{2n}$ combinations for $k$, but this also includes a high level of degeneracy for each resultant wavenumber. Specifically, \cite{mo_paper} showed that the degeneracy of the wavenumber of this circuit is given by $\{{2n\choose2n},{2n\choose2n-1},{2n\choose2n-2},\cdots,{2n\choose0}\}$ for the wavenumbers $\{ -n, -(n-1) , \cdots , -1 , 0 , 1 , \cdots , n - 1, n\}$. Keeping this in mind, one can rewrite Eq.~\ref{eqn:exps_out} as
\begin{eqnarray}
    f(x,\theta) =\sum_{k=-n}^{n} c_k(\theta) e^{ikx} \\= c_0 + \sum_{m=1}^n 2|c_m| cos(m x - arg(c_m)) \\ = c_0 + 2 \sum_{m=1}^n c^r_m cos(mx) - c^i_m sin(mx),\label{eqn:coeffs}
\end{eqnarray}
where $c_m$ are complex coefficients and $(c^r_m,c^i_m) = (Re(c_m),Im(c_m))$.

\subsection{Parallel linear} \label{appendix:par_lin_proof}
A parallel linear encoding consists of a two-qubit system where a Pauli-Z rotation of $x$ is applied to each qubit. This means that two encodings are to be considered: the first encoding considers only the rotation applied to the first qubit making its encoding generator $\mathcal{G}^1 = \sigma_z \otimes \mathbb{I}$ and the second $\mathcal{G}^2 = \mathbb{I} \otimes \sigma_z $:
\begin{equation}
    \mathcal{G}^1 = \frac{1}{2}  \begin{bmatrix}
1 & 0 & 0 & 0 \\
0 & 1 & 0 & 0 \\
0 & 0 & -1 & 0 \\
0 & 0 & 0 & -1 
\end{bmatrix}  ~,~ \mathcal{G}^2 = \frac{1}{2} \begin{bmatrix}
1 & 0 & 0 & 0 \\
0 & -1 & 0 & 0 \\
0 & 0 & 1 & 0 \\
0 & 0 & 0 & -1 
\end{bmatrix}.
\label{eqn:generators}
\end{equation}

Since the gates are applied to separate qubits, the generators commute with each other, and since the group parameters applied to each gate are the same, the elements of the encoding matrix can be written as: 
\begin{equation}
    \left[e^{i \mathcal{G}^1 x}e^{i \mathcal{G}^2 x}\right]_{mn} = e^{i \lambda^1_j x} e^{i \lambda^2_k x} \delta_{mj} \delta_{kn}  = e^{i(\lambda^1_j+\lambda^2_k) x}\delta_{mj}\delta{kn}\\
    = e^{i\mathcal{G}^{\text{comb}}_{mn} x},
\end{equation}
where $\mathcal{G}^{\text{comb}}$ is the combined generator, and $\delta$ is the Kronecker delta which acts as the identity matrix in index notation. Using this expression, one could write
\begin{equation}
    \mathcal{G}^{\text{comb}}_{mn} = \mathcal{G}^1 + \mathcal{G}^2 = \begin{bmatrix}
1 & 0 & 0 & 0 \\
0 & 0 & 0 & 0 \\
0 & 0 & 0 & 0 \\
0 & 0 & 0 & -1 
\end{bmatrix}.
\end{equation}
Similarly to the calculations in App.~\ref{appendix:seq_lin}, this generator provides us with a list of eigenvalues $\lambda \in \{-1,0,1\}$, which, when subtracted from itself yields 2 Fourier frequencies and five wavenumbers $k \in \{-2,-1,0,1,2\}$. This calculation can be generalised to any number of qubits, and Fourier frequencies increase linearly with the number of qubits. 

\section{The derivation for the exponential architectures} \label{appendix:exp_proof}
\subsection{Sequential exponential} 
Starting from Eq.~\ref{eqn:exps_out}, one can scale the sequential features, $x$, by a series of scaling factors $a_l \in \mathbb{N}$ to effectively modify the group generator $\mathcal{G}$ of each layer $l$ to obtain $$e^{-i\mathcal{G}_{i_2,i_1}a_lx} = e^{\lambda_{i_1} a_l x} \delta_{i_1,i_2} = e^{\lambda^{(l)}_{i_1} x} \delta_{i_1,i_2} ,$$ where $\lambda^{(l)}=\frac{1}{2} \{-a_l,+a_l\}$ is the new \textit{effective} list of eigenvalues. For $n$ encoding layers, this work employs:

\begin{equation}
    a_l (l)= 
\begin{cases}
    2^l,& \text{if } l<n-1 \\
    2^l + 1,              & l=n-1
\end{cases},
\end{equation}
where $n-1$ is due to counting from $0$. This parameterisation results in wavenumbers $k_{\boldsymbol{i},\boldsymbol{j}} = (\lambda^{(1)}_{j_1}+\lambda^{(2)}_{j_2}+\cdots)-(\lambda^{(1)}_{i_1}+\lambda^{(2)}_{i_2}+\cdots)$. By choosing the $i$'s and $j$'s such that $\lambda^{(l)}_{j_m}  = {+a_l} \forall m$ and $\lambda^{(l)}_{i_m}  = {-a_l} \forall m$, the largest $k_{\text{max}} = \frac{1}{2}(\sum^{n-1}_{l=0}{a_l} - \sum^{n-1}_{l=0}{-a_l}) = 2^{n}$ is obtained. Symmetrically, the minimal $k$ is $k_{\text{min}} = -2^{n}$, and every whole number between can be created by first flipping from small to large all $\lambda^{(l)}_{j_m}$ to arrive at $k=0$, and then perform the same on $\lambda^{(l)}_{i_m}$ to arrive at the other extreme of the spectrum.  The final list of wavenumbers is an equidistant list of size $2\times2^n +1$, $k_{\boldsymbol{i},\boldsymbol{j}} \in \{-2^n, -2^n+1, \cdots, 0,\cdots, 2^n-1,2^n\}$, which grows exponentially in the number of encodings $n$.

\subsection{Parallel exponential} 
The parallel exponential encoding in the two-qubit limit consists of Pauli-Z rotations of $x$ and $3x$ applied to, respectively, the first and the second qubits. This encoding will use the same generators as Eq.~\ref{eqn:generators} whose purpose is to be multiplied by the rotation angles, $x$ or $3x$ here, and then exponentiated, i.e. $S(x) = e^{-i\mathcal{G}^1 x} e^{-i\mathcal{G}^2 (3x)}$.  $\mathcal{G}^2$ can absorb the scalar $3$ to produce: 

\begin{equation}
    \mathcal{G}^1_{\text{new}} = \frac{1}{2}  \begin{bmatrix}
1 & 0 & 0 & 0 \\
0 & 1 & 0 & 0 \\
0 & 0 & -1 & 0 \\
0 & 0 & 0 & -1 
\end{bmatrix}  ~,~ \mathcal{G}^2_{\text{new}} = \frac{1}{2} \begin{bmatrix}
3 & 0 & 0 & 0 \\
0 & -3 & 0 & 0 \\
0 & 0 & 3 & 0 \\
0 & 0 & 0 & -3
\end{bmatrix},
\end{equation}
where $\mathcal{G}^1_{\text{new}} =\mathcal{G}^1$ remains unchanged.  As the rotation angles are now equal - both are now simply $x$ - one could add the generators to obtain: 
\begin{equation}
    \mathcal{G}^{\text{comb}}_{mn,{\text{new}}} = \mathcal{G}^1_{\text{new}}  + \mathcal{G}^2_{\text{new}}  = \begin{bmatrix}
2 & 0 & 0 & 0 \\
0 & -1 & 0 & 0 \\
0 & 0 & +1& 0 \\
0 & 0 & 0 & -2
\end{bmatrix}.
\end{equation}
This generator produces the eigenvalue list $\lambda \in \{-2,-1,1,2\}$ and by subtracting this list from itself, one can obtain the list of wavenumbers $k \in \{-4,-3,\cdots,3,4\}$, a list of exponential growth with the number of qubits.

\section{Constraints of the sequential exponential}\label{appendix:constrained}
\begin{figure*}[t]
    \centering
    \includegraphics[width=\linewidth]{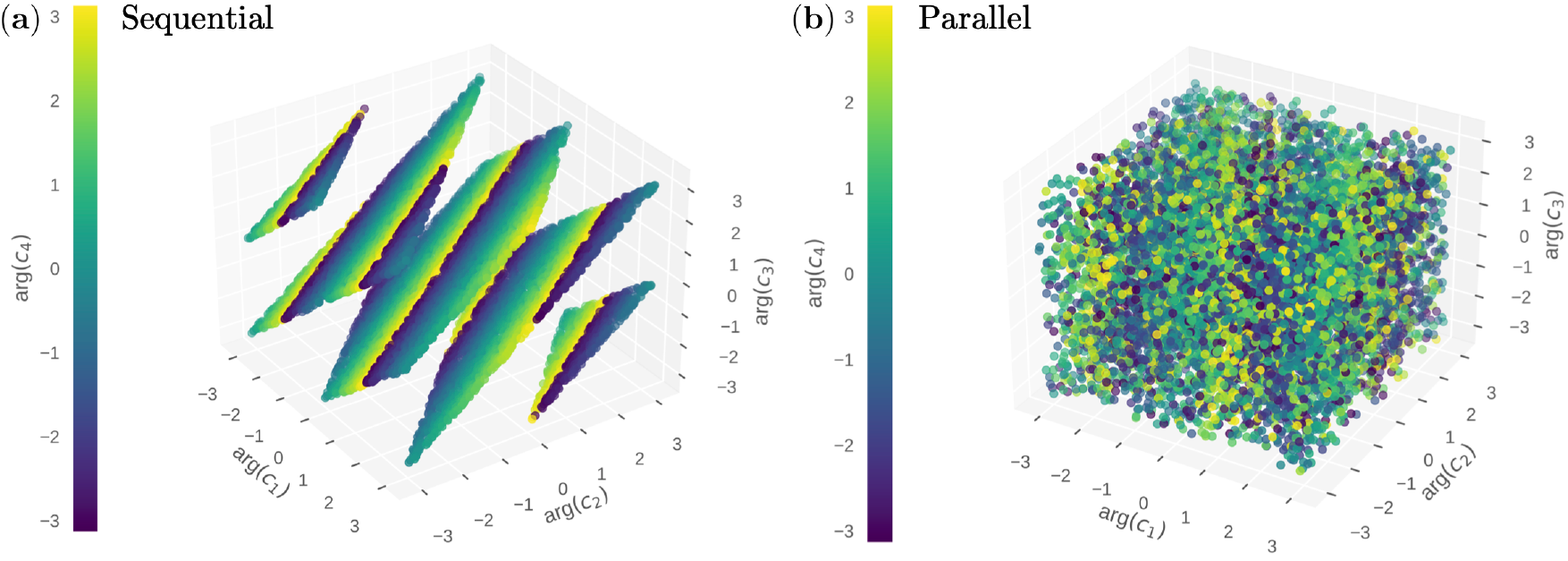}
    \caption{Fourier phases of the two exponential architectures: (a) sequential and (b) parallel. Each architecture was realised $10,000$ times, and their arguments were calculated using the discrete Fourier transform of their outputs. We see that the sequential architecture has a restricted four-dimensional behaviour and that a linear dependence between the phases seems to exist, demonstrating a lack of Fourier accessibility. In contrast, the parallel architecture can fill the space, but still, some constraint is visible between the arg$(c_1)$ and arg$(c_4)$ bases.}
    \label{fig7:constrains}
\end{figure*}

In Sec.~\ref{sec:training}, it was shown that both sequential and parallel exponential architectures represented four Fourier frequencies. However, the latter achieved a lower training loss and a better fit for the top-hat function. This is due to the reduced Fourier accessibility of the sequential architecture, meaning it lacks the freedom to achieve any desired point in the Fourier space. The models with four Fourier frequencies are realised in a nine-dimensional space that includes $c_0$ and the real and imaginary values of $c_i$. Each realisation of the trainable parameters of the quantum circuit produces a 9-dimensional array creating a point in this 9-dimensional space. Realising these two architectures many times makes it possible to analyse the geometry of the Fourier space for each architecture. Still, for manual observation, finding an efficient way to reduce this dimensionality to three dimensions (or four with colour) is essential. Fig.~\ref{fig7:constrains} shows a choice for this dimensionality reduction by investigating the arguments of the complex Fourier coefficients, which, based on Eq.~\ref{eqn:coeffs}, represent the phases of the co-sinusoidal terms. These show that the sequential exponential architecture is dramatically constrained in the collection of phases it can represent and that the parallel exponential is unconstrained in this way.

\section{Beyond exponential growth}\label{appendix:maths_problem}
As in App.~\ref{appendix:seq_lin}, to reach the final list of wavenumbers needed to subtract the eigenvalues of the Hamiltonian in pairs. This section proposes a mathematical problem leading to the highest possible Fourier series.

\begin{problem}
    For a given $m \in \mathbb{N}$, find a list of integers $L \in \mathbb{Z}^m $ such that when subtracted from itself, it produces a new list $k^{(max)}_L = \{x - y~|x,y\in L\}$ whose elements are sequential integers and, except for zero, all the elements have a degeneracy of precisely one. 
\end{problem}

The problem statement produces a list of eigenvalues $L$ from which one can make a list of wavenumbers $k^{(max)}_L$. After finding this list, it is crucial to check if one can create a diagonal Hamiltonian using $R_Z$ rotations and non-parameterised gates whose diagonal elements are the numbers in $L$. In Appendix C.3 of \cite{evanpeters}, this problem is equated to the \textit{perfect Golomb ruler} where for $5\leq m$, this becomes impossible, and the numbers either become nonsequential or degenerate. 

\end{document}